\documentclass[usenatbib]{mn2e}
\usepackage{graphicx}
\usepackage{ifthen}

\def\ltsima{$\; \buildrel < \over \sim \;$}
\def\simlt{\lower.5ex\hbox{\ltsima}}
\def\gtsima{$\; \buildrel > \over \sim \;$}
\def\simgt{\lower.5ex\hbox{\gtsima}}
\def\kms{{\rm\,km\,s^{-1}}}

\def\kpc{{\rm\,kpc}}

\def\msun{{\rm\,M_\odot}}

\def\UseFigs{1}

\def\deg{^\circ}

\def\s{\ifmmode \widetilde \else \~\fi}
\def\={\overline}

\def\spose#1{\hbox to 0pt{#1\hss}}

\def\lta{\mathrel{\spose{\lower 3pt\hbox{$\mathchar"218$}}
     \raise 2.0pt\hbox{$\mathchar"13C$}}}
\def\gta{\mathrel{\spose{\lower 3pt\hbox{$\mathchar"218$}}
     \raise 2.0pt\hbox{$\mathchar"13E$}}}
\def\Dt{\spose{\raise 1.5ex\hbox{\hskip3pt$\mathchar"201$}}}    
\def\dt{\spose{\raise 1.0ex\hbox{\hskip2pt$\mathchar"201$}}}    

\def\dotsfill{\leaders\hbox to 1em{\hss.\hss}\hfill}
\def\sun{\odot}

\def\Gyr{{\rm\,Gyr}}

\loadboldmathitalic 
\title[A dwarf galaxy remnant in Canis Major]
{A dwarf galaxy remnant in Canis Major:\\ 
the fossil of an in-plane accretion onto the Milky Way}
\author[N. F. Martin, R. A. Ibata, M. Bellazzini, M. J. Irwin, G. F. Lewis, W. Dehnen]
{N. F. Martin$^{1}$, R. A. Ibata$^{1}$, M. Bellazzini$^{2}$, M. J. Irwin$^{3}$, 
G. F. Lewis$^{4}$, W. Dehnen$^{5}$\\
$^{1}$
Observatoire de Strasbourg, 11, rue de l'Universit\'e, F-67000, Strasbourg, 
France\\
$^{2}$
INAF - Osservatorio Astronomico di Bologna, Via Ranzani 1, 40127, 
Bologna, Italy\\
$^{3}$
Institute of Astronomy, Madingley Road, Cambridge, CB3 0HA, U.K.\\
$^{4}$
Institute of Astronomy, School of Physics, A29, University of Sydney, NSW
2006, Australia\\
$^{5}$
Department for Physics \& Astronomy, University of Leicester,
Leicester LE1 7RH, U.K.}

\date{\today}
\begin{document} 
\maketitle 
\begin{abstract}
We  present  an analysis  of  the  asymmetries  in the  population  of
Galactic  M-giant  stars  present  in  the 2MASS  All  Sky  catalogue.
Several large-scale asymmetries are  detected, the most significant of
which is a strong elliptical-shaped stellar over-density, close to the
Galactic plane at $(\ell=240^\circ, b=-8^\circ)$, in the constellation
of  Canis Major.   A small  grouping of  globular  clusters (NGC~1851,
NGC~1904, NGC~2298,  and NGC~2808), coincident in  position and radial
velocity, surround  this structure, as  do a number of  open clusters.
The population  of M-giant  stars in this  over-density is  similar in
number to that in the core  of the Sagittarius dwarf galaxy.  We argue
that  this  object  is  the  likely dwarf  galaxy  progenitor  of  the
ring-like structure  that has recently been  found at the  edge of the
Galactic  disk.  A  numerical  study  of the  tidal  disruption of  an
accreted dwarf  galaxy is presented.   The simulated debris  fits well
the  extant  position,  distance   and  velocity  information  on  the
``Galactic  Ring'', as  well as  that of  the  M-giant over-densities,
suggesting that all  these structures are the consequence  of a single
accretion event.   The disrupted dwarf  galaxy stream orbits  close to
the  Galactic Plane,  with  a pericentre  at  approximately the  Solar
circle,  an orbital  eccentricity  similar  to that  of  stars in  the
Galactic thick  disk, as  well as a  vertical scale height  similar to
that of the thick disk.  This finding strongly suggests that the Canis
Major dwarf  galaxy is  a building block  of the Galactic  thick disk,
that the  thick disk  is continually growing,  even up to  the present
time, and  that thick  disk globular clusters  were accreted  onto the
Milky Way from dwarf galaxies in co-planar orbits.
\end{abstract}

\begin{keywords}
Galaxy: structure -- Galaxy: formation -- galaxies: interactions
\end{keywords}

\section{Introduction}

It  is   now  generally  accepted  that  galaxies   assembled  by  the
hierarchical  merging of dark  matter halos  \citep{white78, white91},
with the accretion of luminous (dwarf) galaxies playing a significant,
or  probably  even a  major,  role in  the  formation  of the  visible
structures.   In   the  Milky   Way,  this  picture   is  dramatically
illustrated  by  the existence  of  the  Sagittarius dwarf  spheroidal
galaxy, which  is in  an advanced state  of tidal disruption,  and has
populated the  outer halo with  M giant stars  \citep{ibata02}, carbon
stars \citep{ibata01a} and globular clusters \citep{bellazzini03a}.

The accretion of  the Sagittarius galaxy is currently  the only strong
evidence that the Milky Way is absorbing satellite galaxies.  However,
the ring-like  structure that has recently been  discovered around the
Galaxy  could  be  the  consequence  of  another  such  event.   First
discovered as an  over-density in blue stars of  the Sloan Digital Sky
Survey  \citep{newberg}, this  ``Ring''  has since  been probed  using
photometry  from   the  INT  Wide   Field  Camera  (\citealt{ibata03},
hereafter  I03), SDSS  spectrometry  and photometry  (\citealt{yanny},
hereafter Y03) and 2MASS M-giants (\citealt{rochapinto}).  Surrounding
the   Galactic   disk,  its   Galactocentric   distance  ranges   from
$\sim15\kpc$ to  $\sim20\kpc$ in fields  taken within $30\deg$  of the
Galactic plane and for $122\deg < \ell < 227\deg$.

It has  been proposed (I03  and Y03) that  the structure could  be the
tidal stream  stripped away  by the Milky  Way tides from  a satellite
galaxy that has an orbital plane  close to the plane of the Galaxy. In
light  of this  interpretation,  it is  interesting  to reexamine  the
\citet{abadi} simulation  of the formation of a  disk galaxy assembled
hierarchically in  a $\Lambda{}$CDM  cosmology.  They showed  that the
thin and thick disks of  their simulated galaxy, which were similar to
those of the  Milky Way, do not share the  same origin.  The simulated
thick  disk component  was  constructed mainly  through accretions  of
satellites  with  an  orbit   close  to  the  Galactic  plane.   These
satellites  are dense enough  for their  orbits to  circularize before
they are disrupted.

\citet{helmi03} also  used this simulation to follow  more closely the
behavior of  an in-plane accretion.  They concluded  that it naturally
leads  to the  formation of  a ring-like  structure around  the galaxy
similar  to  the  one  found  by \citet{newberg},  with  two  possible
explanations. One possibility is that the ring is a tidal arc produced
by  stars stripped  away from  the  parent satellite  during a  recent
pericentric passage. This would  produce an asymmetric structure above
and  below  the  disk,  limited  in  Galactic  longitude  and  with  a
significant    velocity   gradient    in   the    Galactic   longitude
direction. Alternatively, it could be a shell-like structure, produced
by ancient minor mergers, similar to the shells observed in elliptical
galaxies.  In this  case, the ring would be  rather symmetric and with
no velocity gradient.

To discriminate between these  competing scenarios, it is essential to
analyze a large-scale, homogeneous data  set.  Until now, this was the
main limitation, but with the release  of the 2 Micron All Sky Survey,
All Sky Release  (2MASS ASR) catalogue we now  have an invaluable tool
to  determine the  characteristics of  this Galactic  ``Ring''.  Apart
from  covering  all the  sky,  the  2MASS  ASR contains  precise  near
infrared photometry, so  it is not much hampered  by extinction in the
low Galactic  latitude regions with high  dust-contamination where the
``Ring'' is primarily located.

We use this  catalogue to probe the ring-like  structure and determine
its dimensions and characteristics on the celestial hemisphere opposed
to the Galactic centre. In section~2,  we discuss the data set we used
for our  study, presented in  section~3. In section~4, we  present the
results  of  our  search  of  clusters  in  the  ring-like  structure.
Section~5 presents N-body simulations of the tidal disruption of the
progenitor  of the  observed  structures. Finally,  in  section 6,  we
discuss  our  results  and  the  consequences on  the  origin  of  the
``Ring''. Throughout this work, we assume that the Solar radius is
$R_\odot = 8\kpc$, that the LSR circular velocity is $220\kms$, and
that the peculiar motion of the Sun is ($U_0=10.00\kms, V_0=5.25\kms,
W_0=7.17 \kms$; \citealt{dehnen98b}).

\section{Data}

The  low Galactic  latitude  regions  that we  study  here are  mainly
hampered by two effects: by reddening due to the high concentration of
dust in the Galactic plane  and by the presence of numerous foreground
disk  stars. The  2MASS near  infrared photometry  has the  ability to
probe the highly  obscured regions, but it cannot  be directly used to
remove the foreground disk stars.  Therefore, in order to diminish the
contamination of the closest disk stars, we cross-identified the 2MASS
and  the  USNO-B1.0  catalogues  in  order  to  complement  the  2MASS
photometry  with proper  motion data.   This was  done for  stars with
$|b|>5\deg$.

In the present study we will  work with two samples of stars: sample A
is simply the 2MASS ASR point-source catalogue (with 2MASS flags ${\tt
bl\_flg=111, cc\_flg=000, gal\_contam=0, mp\_flg=0}$); It will be used
to search for the Galactic ``Ring'' population.  Sample B is identical
to sample A, except that  we have removed all sources with significant
(3$\sigma$)  proper motions.   The contamination  fraction  of objects
with proper  motion corresponds to  $\sim35$\% of the  2MASS sources,
but  since they  are  irregularly distributed  in  a volume  of a  few
kiloparsecs  around the  Sun  this  sample cannot  be  used to  detect
assymetries. The  proper motion discrimination is  however very useful
for   plotting  colour-magnitude   diagrams:  the   disk   dwarfs  are
particularly affected  by this process  and, since they tend  to blend
with red giant populations in CMDs, filtering them allows us to follow
distant  red  giant  branch  (RGB)  populations  up  to  2  magnitudes
deeper. However,  RGB stars beyond  $\sim 5\kpc$ are not  removed with
this  filter,  judging  from  our  tests in  globular  cluster  fields
\citep{martin}.

We de-redden the photometric data using the \citet{schlegel} algorithm
which  interpolates  the  ${\rm E(B-V)}$  value  from the  IRAS  100  micron
emission all-sky  maps. All  the ${\rm J,  H, K_s}$  magnitudes stated
below  are  extinction-corrected.   In   order  to  use  only  precise
photometry, we remove sources with ${\rm K_{s}>14.3}$ (the 10-sigma limit of
the 2MASS  photometry) and  with ${\rm E(B-V)>0.555}$ (which  corresponds to
${\rm E(J-K_{s})>0.3}$).   This last  limit  mainly removes  zones near  the
Galactic plane with $|b|<5\deg$ from our samples.  To make sure we can
compare   quantitatively   the  distribution   of   sources  in   both
hemispheres, we  divide the sky  into $0.1\deg \times  0.1\deg$ pixels
(in Galactic coordinates) flagging as  unusable each zone in which the
extinction exceeds the chosen limit.  However, to maintain North-South
symmetry we also flag as  unusable the symmetric pixel on the opposite
side  of the Galactic  plane.  We  also limit  our present  samples to
$|b|<40\deg$.

\section{Results}

\subsection{M-giant sample}

From sample A, we first select a subsample of late M-giant stars using
the same  colour limits as  \citet{majewski}: ${\rm 0.85 < J-K_{s}  < 1.3}$,
${\rm J-H<0.561(J-K_{s})+0.36}$  and ${\rm J-H>0.561(J-K_{s})+0.22}$.   We measure
the distance of the Galactic M-giants using the RGB of the Sagittarius
dwarf galaxy as a  reference, as calibrated by \citet{majewski}: ${\rm
K_s = -8.650 (J-K_s)_0 +  20.374}$.  By assuming a distance modulus of
${\rm  m-M=16.9}$  for the  stellar  population  at  the core  of  the
Sagittarius galaxy  \citep{ibata97}, one can estimate  the distance to
other RGB  populations.  Figure~1 compares  the 2MASS colour-magnitude
diagrams  (CMDs) of  the Sagittarius  dwarf  with those  of the  Large
Magellanic Cloud  (LMC) and the  globular cluster Omega  Centauri; the
corresponding   distance  calibrations   are   also  displayed.    The
relatively poor  fits for the LMC  and Omega Cen, are  probably due to
differences in age  and metallicity between these objects  and the Sgr
galaxy. Using the Sgr RGB  calibration to measure the distances of the
LMC and  Omega Cen  would incur a  $\sim 30$\% systematic  error, with
these  distances being  underestimated.   This systematic  uncertainty
will cloud all of the distances derived below for the Galactic M-giant
population.

\begin{figure}
\ifthenelse{\UseFigs=1}{
\includegraphics[angle=270,width=\hsize]{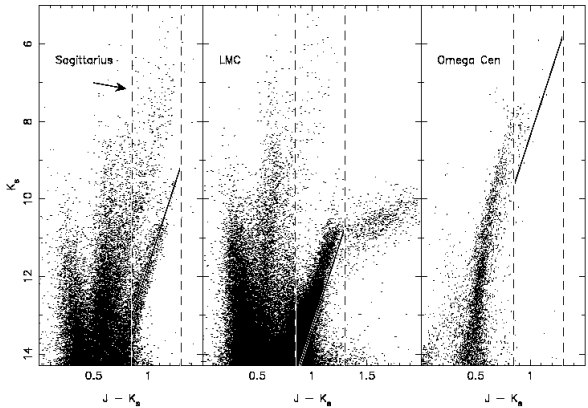}
}{caption1}
\caption{Comparison  of  colour-magnitude   diagrams  of  our  M-giant
distance calibrator,  the Sagittarius dwarf  galaxy (left-hand panel),
with the LMC (middle panel) and Omega Centauri (right-hand panel). The
proper-motion cleaned catalogue (sample B)  is used here.  The Sgr CMD
is derived from  a $1\deg$ circle, centred on the  core of the galaxy,
the LMC CMD  comes from an annulus between  $5\deg$ and $6\deg$, while
the  Omega Cen  CMD  is from  data within  $20'$  of the  core of  the
cluster.   The ${\rm J-K_s}$,  ${\rm K_s}$  colour-magnitude selection
window is  shown between  the dashed vertical  lines (though  note the
additional  selection in ${\rm  J-H}$ colour  is not  displayed here).
The arrow shows  the reddening vector, with a  length corresponding to
the  chosen  reddening  limit  of  ${\rm E(B-V)  <  0.555}$.   In  the
left-hand  panel,  the  solid  line  shows  the  \citet{majewski}  RGB
calibration  for the Sgr  dwarf (${\rm m-M=16.9}$;  \citealt{ibata97}).  The
parallel sequence  approximately two magnitudes brighter  than the Sgr
RGB  is the  RGB  of the  bulge.  The  solid  line in  the middle  and
right-hand  panels  show  the  corresponding  ridge  line  assuming  a
distance modulus  of 18.515 for  the LMC \citep{clementini}  and 13.53
for Omega Cen \citep{harris}.  Using the Sgr fiducial to determine the
distance  to  the  LMC or  Omega  Cen  would  lead  to a  $\sim  30$\%
underestimate of their  distances. The substantial population red-ward
of  ${\rm J  - K_s  = 1.3}$  in the  LMC are  AGB carbon  stars, which
highlights the need for the upper ${\rm J - K_s}$ colour limit.}
\end{figure}

\begin{figure}
\ifthenelse{\UseFigs=1}{
\includegraphics[angle=270,width=\hsize]{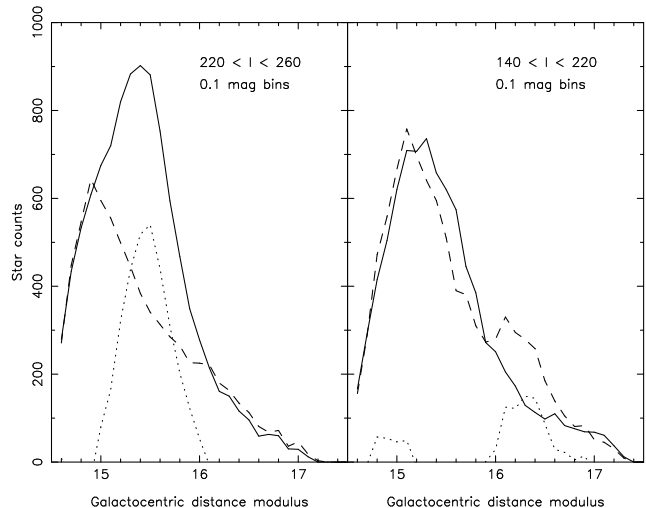}
}{caption1}
\caption{The   Galactocentric  distance  modulus   ${\rm  (m-M)_{GC}}$
distribution for  M-giant stars in  sample A with  $|b|<40\deg$, ${\rm
E(B-V)<0.555}$ and for different ranges in $\ell$.  Note that in this,
and all  subsequent figures, those  regions removed by  the extinction
cut have their corresponding symmetric region about the Galactic plane
removed  as well.  The  star distribution  in the  southern hemisphere
(full  line)  is  compared  to the  northern  hemisphere  distribution
(dashed  line); the  difference  between these  distributions is  also
plotted (dotted line).  The $220\deg<\ell<260\deg$ region (left panel)
shows an  over-density in the southern hemisphere,  ranging over ${\rm
15.0<(m-M)_{GC}<16.0}$ with a  maximum at ${\rm (m-M)_{GC}=15.5}$ (the
signal   to   noise   ratio   of   this  peak   is   $S/N=30$).    The
$140\deg<\ell<220\deg$ region  (right panel) shows  an over-density in
the  northern hemisphere,  ranging  over ${\rm  16.0<(m-M)_{GC}<16.6}$
with   a   maximum   between   ${\rm   (m-M)_{GC}=16.1}$   and   ${\rm
(m-M)_{GC}=16.4}$ (this peak has $S/N=13$).}
\end{figure}

A  comparison between  the northern  and southern  hemispheres  of the
distribution of  source distances, estimated from sample  A, using the
Sgr RGB fiducial described  above, shows two striking asymmetries, one
in each hemisphere (see Figure~2).  The southern over-density, located
between $220\deg <  \ell < 260\deg$ is the  most significant (detected
with  $S/N=30$). It  ranges  from $10.0\kpc$  to  $15.8\kpc$ from  the
Galactic  centre  and  is  centred at  $\sim12.6\kpc$.   The  northern
over-density ranging from  $\ell=140\deg$ to $\ell=220\deg$ is further
away,  between $15.8\kpc$  and  $20.9\kpc$, with  its maximum  between
$16.6\kpc$ and $19.0\kpc$ (this peak is detected at $S/N=13$). In both
cases  there   are  approximatively  twice   as  many  stars   in  the
over-density than in the other hemisphere.

\begin{figure}
\ifthenelse{\UseFigs=1}{
\includegraphics[angle=270,width=\hsize]{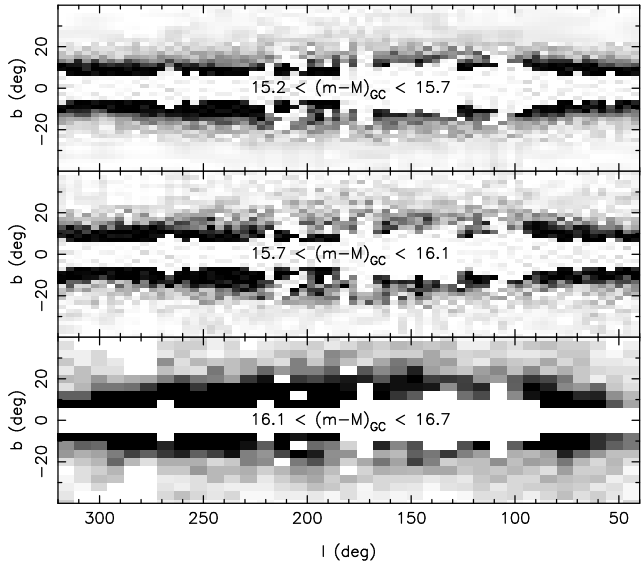}
}{}
\caption{The  distribution  of  M-giant  stars  (from  sample  A)  for
different   ranges   in    Galactocentric   distance   modulus   ${\rm
(m-M)_{GC}}$.    (We  choose   to  display   these   distributions  in
Galactocentric coordinates because  the over-densities are more peaked
than  in Heliocentric  coordinates, whereas  the use  of  the distance
modulus makes  it easier  to visualize the  effects of  extinction and
population differences).  In the  upper two panels a pixel corresponds
to a $4\deg \times 2\deg$ rectangle, while in the bottom panel a pixel
covers $8\deg  \times 4\deg$; the  scale colour is different  for each
panel to  emphasize the structures.   The circular hole at  $\ell \sim
280\deg$  and $b \sim  -35\deg$ comes  from the  removal of  the Large
Magellanic Cloud; due to the symmetry imposed on the map, this hole is
reflected in the North as well.  The top panel contains the sources of
our  sample with ${\rm  15.2<(m-M)_{GC}<15.7}$.  The  most significant
over-density in  this distance range is  created by a  dense object at
$220\deg\lta \ell  \lta 260\deg$ and  $b \gta -20\deg$,  towards Canis
Major.  The middle panel shows  the distribution of sources with ${\rm
15.7<(m-M)_{GC}<16.1}$,   while   the   bottom   panel   shows   ${\rm
16.1<(m-M)_{GC}<16.7}$. In the top,  middle and bottom panels, a black
pixel corresponds to 40, 16 and 16 counts, respectively.}
\end{figure}

\begin{figure}
\ifthenelse{\UseFigs=1}{
\includegraphics[angle=270,width=\hsize]{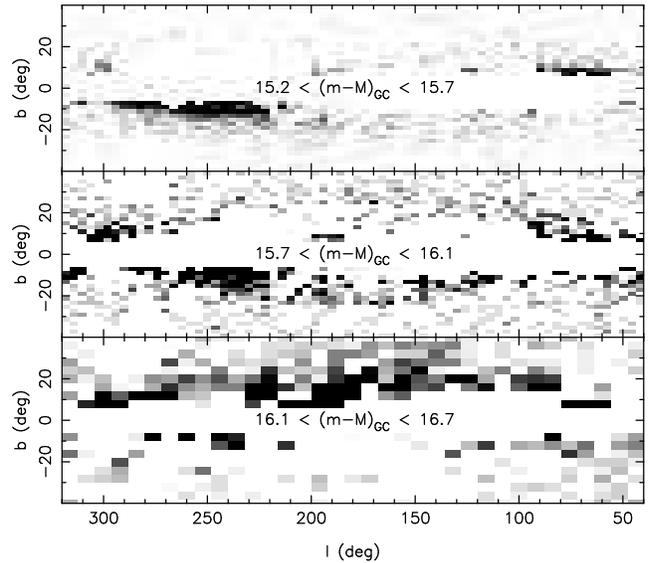}
}{}
\caption{As Figure~3, but showing the difference in M-giant starcounts
between the  two hemispheres.  The Canis Major  over-density at $(\ell
\sim 240^\circ, b \sim -10^\circ)$ is clearly visible on the upper two
panels.  The middle panel also  shows a fainter population arcing over
the  sky  between $110\deg\lta  \ell  \lta  210\deg$  in the  southern
hemisphere.   Another  huge  arc-like  structure  is  present  in  the
northern  hemisphere  in  the  bottom  panel  ranging  from  $\ell\sim
140\deg$ to $\ell\sim 220\deg$.  In the top, middle and bottom panels,
a black pixel corresponds to 40, 8 and 8 counts, respectively.}
\end{figure}

\begin{figure}
\ifthenelse{\UseFigs=1}{
\includegraphics[bb= -105 70 640 725,clip,width=\hsize]{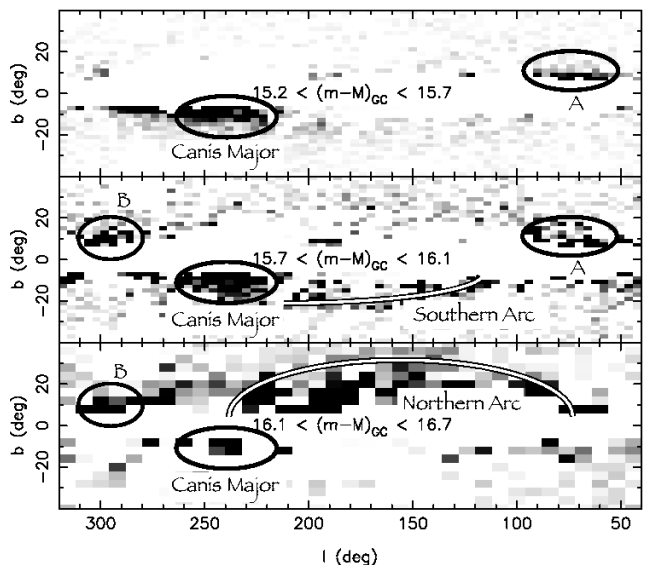}
}{}
\caption{As Figure~4, with  an overlay of the main  asymmetries in the
Galactic  M-giant  distribution  detected  in Figures~3  and  4.   The
strongest of  these is the  Canis Major over-density, followed  by the
Northern  Arc. A  fainter Southern  Arc is  also visible,  as  are two
structures  at  $\ell=70\deg$,  and  $\ell=300\deg$,  which  we  label
structures A and B respectively.}
\end{figure}

To examine the structures  creating these two over-densities, the data
in  sample A  was  binned  in Galactic  coordinates,  as presented  in
Figure~3.  In those  pixels where more than 50\% of  the area is below
the  extinction limit,  we  have  corrected the  counts  for the  area
incompleteness; otherwise, the pixel values are set to zero. We impose
North-South  symmetry to  be able  to compare  the counts  between the
hemispheres.    In   the    southern   hemisphere,   for   the   ${\rm
15.2<(m-M)_{GC}<15.7}$    range,    an    object   is    centred    at
$\ell\sim240\deg$  and $b\sim-8\deg$,  in the  direction of  the Canis
Major constellation.   It is  very dense  at its core  --- with  up to
eight times the number of stars  compared to the symmetric zone in the
northern hemisphere  --- and it seems to  be approximately elliptical.
These  structures  are  more  clearly  seen in  the  difference  image
displayed in  Figure~4, where we have subtracted  the star-counts from
the opposite hemisphere.  The Canis Major over-density is striking, as
is  the  wide arc-like  structure  in  the  northern hemisphere.   The
location of these features is  made clearer with schematic overlays in
Figure~5.

To measure  the dimensions of the  Canis Major object,  we examine the
profiles  in   the  latitude  and   Heliocentric  distance  directions
(Figure~6).  Analysis  of the vertical distribution  of the population
is  hampered   by  the  lack  of   non-dust-contaminated  regions  for
$|b|<7\deg$. This  puts a  high degree of  uncertainty on  the profile
fits since the population  could correspond to a Gaussian distribution
centred  at  $b=-6.3\deg\pm1.0\deg$ with  a  FWHM  vertical height  of
$1.6\pm0.2\kpc$, or alternatively,  an exponential distribution with a
$0.73\pm0.05\kpc$ scale  height.  It is interesting to  note that this
scale height derived for the Canis Major structure at $\ell=240^\circ$
is statistically  identical to that  obtained by I03 for  the Galactic
Ring, in a field at  $\ell=123^\circ$.  This lends some support to the
possibility  that  the   structures  are  related.   The  Heliocentric
distribution  of  the object  can  be fit  by  a  Gaussian centred  at
$D_{\sun}=7.1\pm0.1\kpc$  with  FWHM  of $4.2\pm0.3\kpc$.   While  the
centre of the  distribution can be measured reliably,  its FWHM should
be  taken with  caution  since  the distribution  is  skewed to  large
distances.   This could  be due  to the  angle of  observation  of the
structure if it is elongated  in a direction that is not perpendicular
to the  line of sight.   The precise extent  of this object  along the
longitude  axis is  harder  to compute  because  the observations  are
hampered by high  extinction regions and by the  rapid increase in the
number of disk stars at low latitude.

\begin{figure}
\ifthenelse{\UseFigs=1}{
\includegraphics[angle=270,width=\hsize]{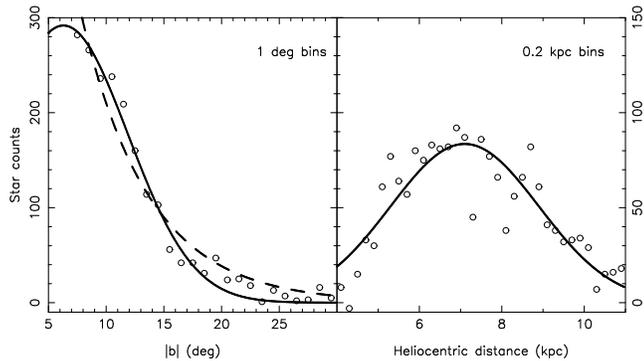}
}{}
\caption{Vertical  and  radial profiles  of  the  Canis Major  object,
derived from sample A.  The northern hemisphere is used as a reference
and  its  star  counts  are  subtracted from  those  of  the  southern
hemisphere.  The  left-hand panel  shows the latitude  distribution of
stars with ${\rm 15.2<(m-M)_{GC}<15.7}$ and $220\deg < \ell < 260\deg$.  The
lack of data for $|b|<7\deg$ puts  a high degree of uncertainty on the
fits,  but   the  best  Gaussian   fit  (full  line)  is   centred  on
$b=-6.3\deg\pm1.0\deg$   with    a   FWHM   of   $13.1\deg\pm1.3\deg$,
corresponding  to  a  latitude  width  of  $1.6\pm0.2\kpc$.   We  also
calculated the  best exponential fit of the  population (dashed line):
it  has a  scale height  of $0.73\pm0.05\kpc$.   The  right-hand panel
shows the radial distribution of stars for the region of the centre of
the object  ($230\deg < \ell < 250\deg$  and $-12\deg<b<-7\deg$).  The
best  Gaussian  fit  has  been  superposed  on  the  star  counts:  it
corresponds to a Gaussian  centred on $D_{\sun}=7.1\pm1.3\kpc$, with a
$4.2\pm0.3\kpc$ FWHM.  We  used the Heliocentric distance distribution
here because our angle selection was made in Heliocentric coordinates,
more suited for observations along these lines of sight.}
\end{figure}

The    Northern    Arc    over-density    is   visible    for    ${\rm
  16.1<(m-M)_{GC}<16.7}$ (bottom panel of Figures~3 and 4).  It begins
near  the Galactic  plane at  $\ell\sim230\deg$  and then  arcs up  to
$\ell\sim140\deg$ and $b\sim+30\deg$.  The  best region to compute its
vertical  height  is  between  $140\deg<\ell<180\deg$  because  it  is
further away  from the  Galactic disk than  elsewhere.  We  find there
that the Northern  Arc has a Gaussian profile  (see Figure~7), centred
on  $19.9\deg\pm1.0\deg$ and  with a  $2.2\pm0.4\kpc$  vertical width.
However, high extinction regions could be responsible for the decrease
of star  counts under  $b\sim15\deg$ and the  vertical width  could be
underestimated.   The thickness  of the  structure was  also computed,
this time in  the region where the over-density  is most pronounced in
the northern  hemisphere, that is for  $140\deg < \ell  < 180\deg$ and
$+10\deg<b<+25\deg$.  The right-hand panel  of Figure~7 shows the best
Gaussian  fit for  the radial  profile  of the  structure: centred  on
$D_{GC}=18.1\pm0.1\kpc$,  it has  a FWHM  line of  sight  thickness of
$2.2\pm0.3\kpc$.   This  structure  is  almost certainly  due  to  the
M-giants present in  the ``Galactic Ring'' of I03  and Y03.  Even with
the rather noisy values derived here, the Gaussian vertical FWHM falls
in the range of the  Y03 and I03 exponential scale height measurements
(respectively $< 4\kpc$ and $\sim2\kpc$).

\begin{figure}
\ifthenelse{\UseFigs=1}{
\includegraphics[angle=270,width=\hsize]{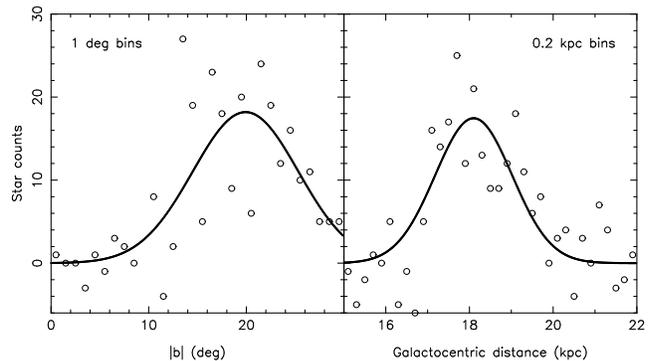}
}{}
\caption{Profiles  of   the  Northern  Arc   structure.  The  southern
hemisphere is used  as a reference and its  star counts are subtracted
from those of the northern  hemisphere.  The left-hand panel shows the
latitude profile of  the structure for $140\deg <  \ell < 180\deg$ and
${\rm 16.1<(m-M)_{GC}<16.7}$,  and  the  best   Gaussian  fit  of  the  star
counts. The distribution has  a mean value of $19.9\deg\pm1.0\deg$ and
a FWHM  of $12.7\deg\pm2.3\deg$. At  the Heliocentric distance  of the
object  ($\sim10\kpc$),   it  corresponds  to  a   vertical  width  of
$2.2\pm0.4\kpc$.  The right-hand panel shows the radial profile of the
structure for $140\deg < \ell < 180\deg$ and $+10\deg<b<+25\deg$.  The
distribution   is   centred    on   $D_{GC}=18.1\pm0.1\kpc$   with   a
$2.2\pm0.3\kpc$  thickness.  Since  here the  observations are  in the
direction  of  the anticentre,  we  were  able  to use  Galactocentric
distances.}
\end{figure}

A  second southern  over-density,  the ``Southern  Arc'' in  Figure~5,
ranges over $110\deg < \ell < 210\deg $ and $12.0\kpc<D_{GC}<15.1\kpc$
(i.e.   ${\rm 15.4<(m-M)_{GC}<15.9}$). This  structure is  detected at
$S/N=8$.  It is possible that  its vertical distribution has a similar
Gaussian profile  to that  of the Northern  Arc, but the  large highly
dust-contaminated region  near the disk hampers  the interpretation of
the  observations.   This  feature  may  correspond  to  the  southern
hemisphere population reported by  Y03 in the longitude range $180\deg
< \ell < 227\deg$ and  at a distance of $R_{GC}=20\pm2\kpc$.  The 30\%
systematic  distance error  mentioned earlier  may be  responsible for
the difference in these two distance estimates.

\begin{figure}
\ifthenelse{\UseFigs=1}{
\includegraphics[angle=270,width=\hsize]{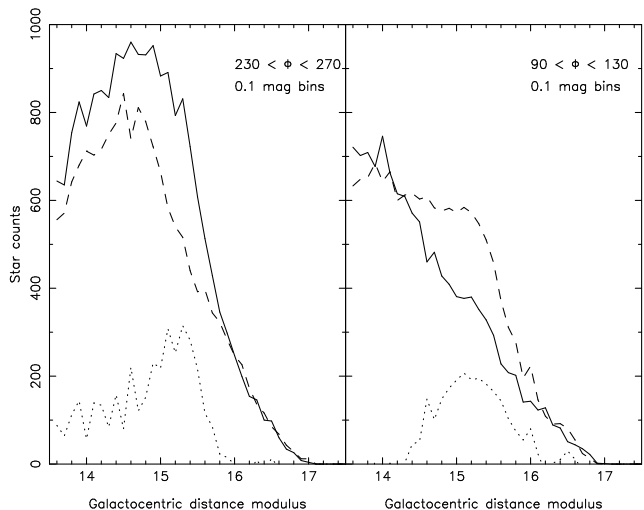}
}{}
\caption{The   Galactocentric  distance  modulus   ${\rm  (m-M)_{GC}}$
distribution for M-giant stars with $|b|<30\deg$, ${\rm E(B-V)<0.555}$
and for different ranges  of the Galactocentric cylindrical coordinate
$\Phi$. The  star distribution in the southern  hemisphere (full line)
is compared to the northern hemisphere distribution (dashed line); the
over-densities  are also  plotted (dotted  line).  The  $230\deg< \Phi
<270\deg$  region  (left  panel),  which  corresponds  to  $\ell  \gta
270\deg$,  shows the  prolongation  of the  Canis Major  over-density.
With a Galactocentric distance  modulus of ${\rm (m-M)_{GC}<15.8}$, it
is closer to the Galactic centre than the over-density we presented on
Figure~2;  the  ${\rm (m-M)_{GC}=13.5}$  lower  limit  is  due to  the
presence of the  bulge.  The over-density has a  maximum star count of
$\sim300$  (per  interval  of   0.1  in  distance  modulus)  at  ${\rm
(m-M)_{GC}=15.1}$ and  slowly decreases with  decreasing distance from
the  Galactic  centre.   The  $90\deg< \Phi  <130\deg$  region  (right
panel),  which corresponds to  $\ell \lta  90\deg$, shows  the feature
labeled as structure ``A'' in  Figure~5, which may be the prolongation
of  the  Northern  Arc.   It   is  also  nearer,  ranging  over  ${\rm
14.3<(m-M)_{GC}<16.1}$ centred around  ${\rm (m-M)_{GC}=15.1}$, with a
profile that is approximately symmetric.}
\end{figure}

The two main  over-densities can be followed around  half of the Milky
Way. To diminish the projection problems that appear when observing in
the quadrant towards the Galactic centre (i.e.  for $\ell>270\deg$ and
$\ell<90\deg$),   we  switched   to  the   Galactocentric  cylindrical
coordinates $(R,\Phi,z)$.  We also  limited our sample to $|b|<30\deg$
since the  sources are further  away from the  Sun. The left  panel of
Figure~8  shows the  Galactocentric distance  modulus  distribution of
stars for  $230\deg <  \Phi < 270\deg$  where the prolongation  of the
Canis Major over-density is visible  despite the higher number of disk
stars.    It  ranges   up  to   $D_{GC}\sim14.5\kpc$  from   at  least
$D_{GC}=5\kpc$   but  the   presence   of  the   bulge  prevents   the
determination of a precise lower  limit. The high radial extent of the
over-density is certainly due to the angle of observation.  Indeed, it
is possible  that the structure  is, at least partially,  not observed
perpendicularly but  rather tangentially. This would  also explain its
broad profile and seems to  be confirmed when taking smaller ranges of
$\Phi$.  The right-hand panel corresponds to $90\deg < \Phi < 130\deg$
and  shows the  prolongation  of  the Northern  Arc.   It ranges  over
$7.2\kpc \lta D_{GC} \lta 16.6 \kpc$ which is nearer from the Galactic
centre  than  it  was in  the  direction  of  the anticentre.   It  is
surprising to find  that it is also denser. This could  also be due to
the angle of  observation: if here too we  are partially observing the
structure tangentially,  the star counts  will be summed up  along the
line of sight.

To summarize, we were able to detect two substantial over-densities of
M-giants,  the Canis Major  over-density and  the Northern  Arc, which
together  surround half of  the Milky  Way, at  varying Galactocentric
radius.  This population can be  as close as $5\kpc$ from the Galactic
centre for  $\Phi=90\deg$ and  $20\kpc$ away in  the direction  of the
anticentre. We also find it  probable that it passes from the northern
to the southern hemisphere at around $\ell=230\deg$.

\subsection{Colour-Magnitude Diagrams}

We now examine the stellar  populations of the stars in the structure,
using  the near  infrared 2MASS  photometry.  As  can be  discerned in
Figure~9, the  ${\rm (J-K_{s},K_{s})}$ CMD of a $5\deg  \times 2\deg$ region
at the centre of the Canis Major object shows two striking features: a
red-giant branch population which does not appear on the reference CMD
on the other side of the Galactic plane and a high number of stars for
${\rm 0.55  \lta J-K_{s}  \lta 0.65}$  and ${\rm 12  \lta K_{s}  \lta  13.5}$. For
comparison, we  superimpose on Figure~9  the line of the  fiducial RGB
sequence of the Sgr dwarf,  adjusted to $7.1\kpc$, the distance of the
Canis  Major  structure  determined  from the  analysis  of  Figure~6.
Inspection  of the  subtracted  Hess-diagram shown  in the  right-hand
panel of  Figure~9 reveals that  the population in this  central field
$5\deg \times  2\deg$ is slightly ($\sim 1.5\kpc$)  more distant, than
estimated above. An  alternative explanation could be that  there is a
stellar population gradient towards the centre of the structure.

This excess RGB  population does not appear to  be spatially separated
from the  Galactic disk,  that is,  we do not  detect a  magnitude gap
separating  this  population and  the  disk  red  giants in  the  data
presented in the left-hand  panel of Figure~9. The structure therefore
lies within, or at the edge,  of the Galactic disk.  The other feature
that is present at the bottom  of this red giant branch (at ${\rm 0.55
<  J-K_s < 0.65}$,  ${\rm K_s  \sim 12.5}$)  is the  corresponding red
clump of the stellar population.   An additional blue plume at ${\rm J
- K_s <  0.5}$ is  seen in the  subtracted Hess diagram;  this feature
could  be an  intermediate-age main-sequence  population of  the Canis
Major galaxy,  although it could possibly  also be an  artifact of the
subtraction.   A deeper optical  wide-field survey  of the  region may
help resolve this issue.

\begin{figure}
\ifthenelse{\UseFigs=1}{
\includegraphics[angle=270,width=\hsize]{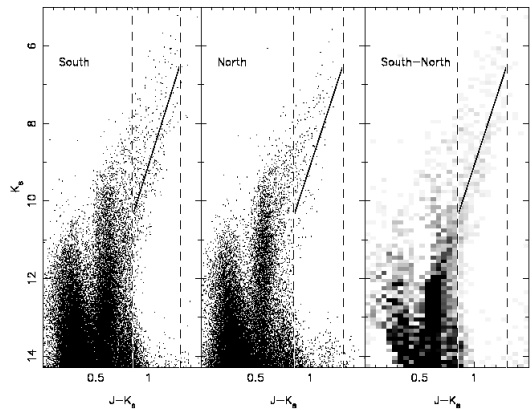}
}{}
\caption{Near    infrared   colour-magnitude   diagrams    (from   the
proper-motion  cleaned sample  B) of  the  centre of  the Canis  Major
object  ($239\deg < \ell  < 244\deg$  and $-10\deg<b<-8\deg$)  and its
symmetric  region  in  the  northern  hemisphere ($239\deg  <  \ell  <
244\deg$ and $8\deg<b<10\deg$).  The  Canis Major object CMD shows two
features which are not in the reference CMD, and that are clearly seen
in the subtracted Hess diagram  shown on the right-hand panel (a black
pixel  corresponds to a  count of  25 stars).   A red-giant  branch is
present from (${\rm  J-K_s \sim 0.7}$, ${\rm K_s  \sim 12}$) to (${\rm
J-K_s  \sim 1.2}$, ${\rm  K_s \sim  8}$), while  the dense  feature at
(${\rm 0.55 < J-K_s < 0.65}$, ${\rm K_s \sim 12.5}$) is the red clump.
As in  Figure~1, the  dashed lines show  part of  the colour-magnitude
selection region.   The solid lines show  the RGB fiducial  of the Sgr
dwarf galaxy (left-hand panel of  Figure~1), adjusted to a distance of
$7.1\kpc$ (the distance to  the structure determined from the analysis
presented in Figure~6).}
\end{figure}

The colour-magnitude diagram of the  Northern Arc reveals a sparse red
giant   branch,   consistent  with   a   Galactocentric  distance   of
$\sim18.1\kpc$, as  we show in  the left-hand panel of  Figure~10.  It
appears that there is a small  gap between the RGB of this population,
and the  disk RGB  stars, indicating that  the structure  is separated
from the disk at its most distant point. The middle panel of Figure~10
shows a comparison region below  the Galactic plane; here, we may also
have  detected a  similar RGB  population, though  the signal  is very
weak. This  feature may correspond  to the ``Southern  Arc'' structure
highlighted in Figure~5.

\begin{figure}
\ifthenelse{\UseFigs=1}{
\includegraphics[angle=270,width=\hsize]{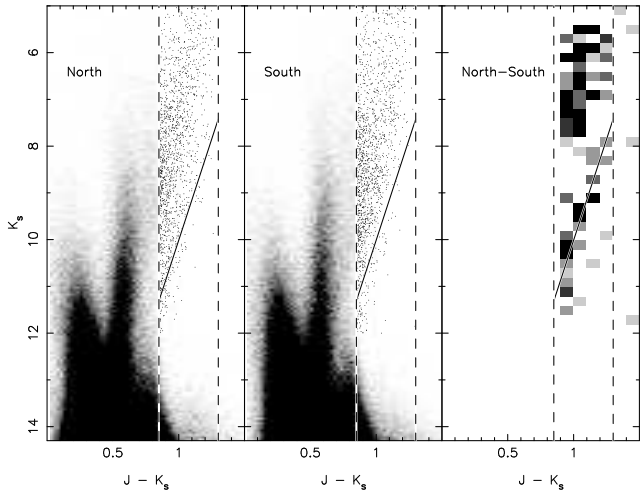}
}{}
\caption{The left-hand panel  shows the near infrared colour-magnitude
diagram (from the proper-motion cleaned  sample B) of a $15\deg \times
15\deg$ region  in the  Northern Arc over-density  ($140\deg <  \ell <
155\deg$  and $10\deg  <  b <  25\deg$).   Since the  Northern Arc  is
fainter  than the  Canis Major  structure, it  is necessary  to gather
sources over  a wider area to  obtain sufficient signal  to render the
excess M-giant population visible. For clarity, we use a Hess diagram,
except in the colour-magnitude region  ${\rm J-K_s > 0.85}$, ${\rm K_s
< 12.0}$.  The solid line shows the position of the fiducial Sgr dwarf
RGB, shifted  to an equivalent Galactocentric  distance of $18.1\kpc$,
as indicated by  the analysis presented in Figure~7.   A population of
stars is clumped  around this line, indicating the  presence of an RGB
at  this  distance.   A  gap  between  the RGB  of  the  Northern  Arc
over-density  and  the  RGBs  of  the  disk  can  also  be  perceived,
indicating  that the  structure is  separated  from the  disk in  this
direction.  The  middle panel  shows the Hess  diagram and CMD  of the
symmetric region about the Galactic  Plane ($140\deg < \ell < 155\deg$
and $-25\deg < b < -10\deg$).  A hint of the same RGB is also present,
though the  population is much less numerous;  it probably corresponds
to  the ``Southern  Arc''  identified  in Figure~5.   Due  to the  low
contrast  of the population  over the  numerous foreground,  it proved
hard to construct  a clear subtracted Hess diagram.  Our best attempt,
shown on  the right-hand  panel, reveals the  presence of  an RGB-like
population beneath the solid line; here a black pixel corresponds to 5
stars. The red clump population, however,  is lost in the noise of the
local disk red dwarf sequence.}
\end{figure}

We also searched for the RGB  population of the structure in small low
extinction windows at  very low latitude.  However, the  low number of
the M-giants  of interest compared to the high number  of stars in the
disk  prevents any further  conclusion on  the connection  between the
Northern Arc and the  Canis Major over-densities.  The RGB populations
for $230\deg < \Phi < 270\deg$ and $90\deg < \Phi < 130\deg$ also have
low contrast  because the structures are  mixed with the  disk and the
higher  number   of  disk  stars   weakens  the  signal  due   to  the
over-densities.

\subsection{Extinction}

Could these  inferred over-densities  simply be due  to errors  in the
\citet{schlegel}  extinction maps?   The  colour of  the vertical  CMD
feature of  disk dwarfs in the extinction-corrected  samples is almost
identical (to  within $\sim  0.1$ magnitudes) North  and South  of the
Galactic plane.  This is  demonstrated in Figure~9, where the sequence
of disk dwarfs is seen both on the left-hand panel at ${\rm J-K_s \sim
0.6}$ in  the over-density  at $(\ell=242^\circ, b=-9^\circ)$,  and at
the  same colour  in  the  symmetric field  about  the Galactic  Plane
$(\ell=242^\circ,   b=9^\circ)$.   Additionally,   this   sequence  is
detected in low  extinction regions at the same  ${\rm J-K_s}$ colour,
giving us further confidence that the \citet{schlegel} extinction maps
are not significantly in error along these sight-lines.

For the RGB-like population present  on the subtracted Hess diagram on
the right-hand  panel in Figure~9  to be caused by  extinction errors,
would require  a contaminating population  that runs parallel  to that
RGB  population (but  both bluer  and brighter).   The  only available
population that  has this  property is the  RGB of the  Galactic disk.
However, inspection of Figures~2 and  8 shows that the distribution of
M-giants on  opposite sides  of the Galactic  Plane match well  at the
lower  distance limits  and  at  the upper  distance  limits of  those
diagrams.   It is  at  intermediate distances  that  a discrepancy  is
detected.    Extinction  errors  can   only  rearrange   the  distance
distributions, they  cannot cause  an apparent overabundance  of stars
all along the RGB. For these  reasons it is unlikely that the detected
over-densities are extinction artifacts.

\subsection{Mass of the Canis Major structure}

To estimate the mass of the  Canis Major object, we compare the number
of  detected  M-giants to  those  observed  in  the Sagittarius  dwarf
galaxy. For Canis Major, we  subtract the northern hemisphere sample A
data from the southern hemisphere  and sum under the profiles shown in
Figure~6.   Within a $10\deg$  radius region  (but which  misses $\sim
40$\% of the  area due to the $|b|<5\deg$ cutoff and  due to the ${\rm
E(B-V) < 0.555}$  reddening limit) we detect 2300  M-giants.  The same
measurement around the  core of the Sagittarius dwarf  galaxy gives an
almost identical  number of 2200  M-giants.  With the caveat  that the
stellar populations  are not  the same, this  suggests that  the Canis
Major  object  has a  similar  total luminosity  to  the  core of  the
Sagittarius  dwarf galaxy,  which has  been measured  to be  ${\rm M_V  = -
13.27}$ \citep{majewski}. If  we further assume that the  mass to light
ratio of the two galaxies is similar, we conclude that the Canis Major
object  has an  almost  identical mass  to  that of  the  core of  the
Sagittarius dwarf.  The mass of the  latter is a matter of some debate
\citep{ibata97,  ibata98,   helmi01,  gomez},  but   a  range  between
$10^8\msun$ to $10^9\msun$ seems plausible, with the upper bound being
the likely pre-disruption mass of that galaxy.

\section{A clustering of globular clusters}

While searching for structures in the Galactic globular cluster system
that  could be  connected with  the stream  of the  Sagittarius galaxy
\citep{bellazzini03a,bellazzini03b}, we noted  an interesting group of
clusters  that  are not  related  with  Sgr  and that  are  remarkably
confined  in phase space  \citep{bellazzini03c}. These  four clusters,
NGC~1851,  NGC~1904, NGC~2298  and NGC~2808,  are quite  close  to one
another in space, indeed, they may be enclosed in a sphere with radius
$R\simeq   6\kpc$   centred    at   Galactic   Cartesian   coordinates
$(X,Y,Z)=(-11.5,-8.9,-4.5)  \kpc$.   Among  the  outer  halo  globular
clusters \cite[those with $R_{GC}>10\kpc$, see][]{bellazzini03a} there
is only  one other  group of globulars  that shows  similar clustering
properties, the clusters associated with the main body of the Sgr dSph
galaxy  (M~54, Ter~7,  Ter~8, and  Arp~2).  In  Figure~11  the spatial
distributions  of  the  two  groups  are  compared  in  two  different
projections   of  Cartesian   Galactocentric   coordinates:  NGC~1851,
NGC~1904, NGC~2298 and NGC~2808 are as close one to another as the Sgr
globulars. It is  interesting to recall that the  similar proximity of
the four Sgr cluster was not noted until the discovery of their parent
galaxy \citep{bellazzini03c}.

\begin{figure}
\ifthenelse{\UseFigs=1}{
\includegraphics[angle=0,width=\hsize]{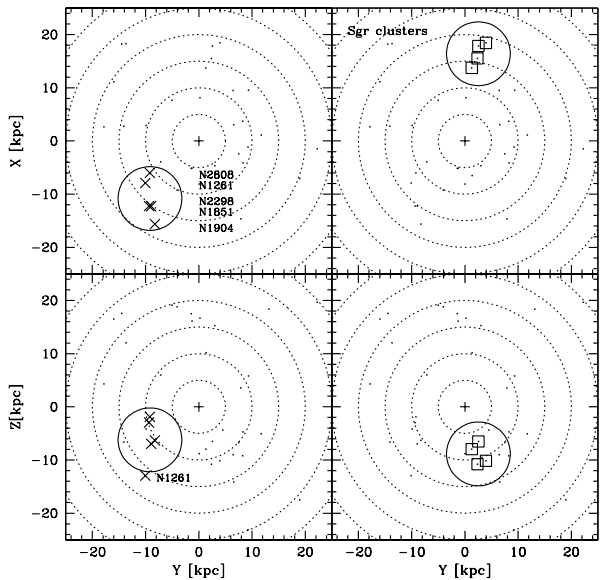}
}{}
\caption{Comparison of the clustering properties in space of the group
of  globulars NGC~1851, NGC~1904,  NGC~2298, NGC~2808,  NGC~1261 (left
panels) and of the globular  clusters associated with the main body of
the Sgr  galaxy (right  panels; M~54, Ter~7,  Ter~8, and  Arp~2).  The
clusters  are plotted in  Cartesian Galactocentric  projections (upper
panels: YX; lower panels: YZ).  In this coordinate system the Galactic
Centre   is    at   $(X,Y,Z)=(0,0,0)\kpc$   and   the    Sun   is   at
$(X,Y,Z)=(-8,0,0)\kpc$.  The small cross indicates the Galactic Centre
and the  dotted circles have  Galactocentric radius from  $R=5\kpc$ to
$R=30\kpc$,  in steps  of  $5\kpc$.   Both groups  are  enclosed in  a
(continuous) circle having radius $r=6\kpc$.}
\end{figure}

If we consider the radial velocities, the present Sgr globular cluster
system appears  much more compact  ($\sigma_{V_{LSR}}=21.5 \kms$) than
the considered group  ($\sigma_{V_{LSR}}=95.3 \kms$). However the data
presented in Figure~12 show  that the LSR-corrected radial velocity of
the group members correlates  well with Galactic longitude.  Moreover,
their   Galactocentric   distance   also  correlates   with   Galactic
longitude. These  facts are strongly  suggestive of a  coherent motion
along a common orbit.

After  the  discovery  of  this  remarkable phase  space  grouping  of
globular  clusters  we  noted   that  their  positions  are  virtually
coincident with the Canis Major structure discussed in this paper. The
strong analogy with the Sgr case  brings us to the conclusion that the
two structures  are associated  and that NGC~1851,  NGC~1904, NGC~2298
and  NGC~2808 are  likely the  local remnant  of the  globular cluster
system of the galaxy that was the progenitor of the Canis Major object
and the Northern Arc.

These  clusters  also  show  some interesting  similarities  in  their
intrinsic properties.  For instance,  their metallicity is enclosed in
the relatively limited  range ${\rm -1.03 \le [Fe/H]  \le -1.71}$, all
of   them  show   an  extended   blue  Horizontal   Branch   in  their
colour-magnitude diagrams, and NGC~1851  and NGC~2808 are the only two
halo clusters showing simultaneously a  well populated red clump and a
blue   tail   in  their   HB   morphology  \cite[see][and   references
therein]{harris,bedin,bellazzini}.

There are  other globular  clusters that may  be related to  the Canis
Major  structure.  NGC~1261  is  also  nearby in  space  to the  above
described group but  its velocity does not fit  well with the velocity
gradients  shown  in  Figure~12. We  will return  to  this  issue  in
Section~5 below.

\begin{figure}
\ifthenelse{\UseFigs=1}{
\includegraphics[angle=0,width=\hsize]{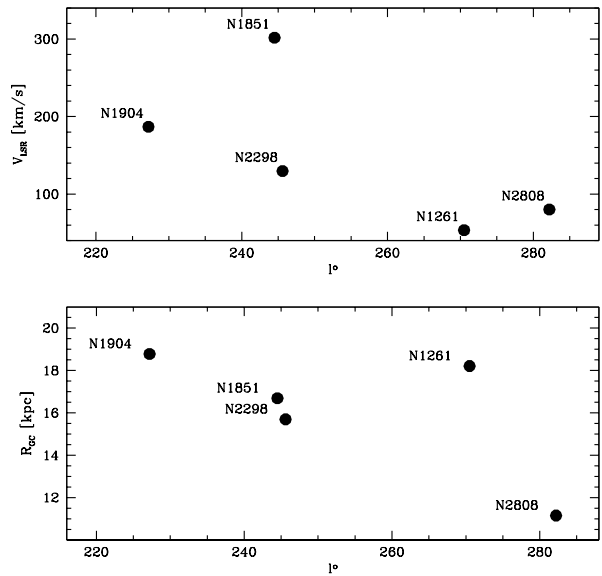}
}{}
\caption{The  upper  panel  shows  the  correlation  between  Galactic
longitude and  the radial  velocity (corrected for  the motion  of the
Local Standard  of Rest)  for the clusters  associated with  the Canis
Major  object.  The  lower  panel  presents  the  correlation  between
Galactic longitude and Galactocentric distance.}
\end{figure}

\begin{figure}
\ifthenelse{\UseFigs=1}{
\includegraphics[angle=270,width=\hsize]{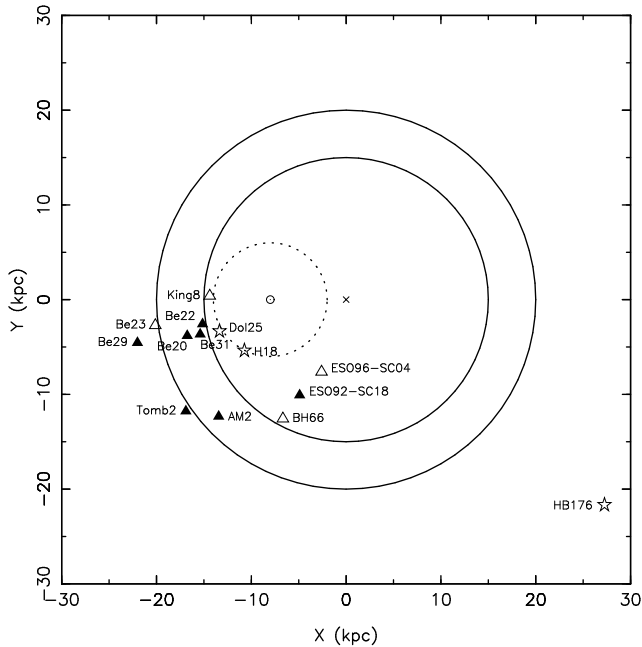}
}{}
\caption{Distribution  of open  clusters  with $D_{\sun}>6\kpc$.   The
Galactic centre  has been  marked (cross) as  well as the  Sun (dotted
circle). The  full line circles  correspond to a distance  of $15\kpc$
and $20\kpc$  from the Galactic centre; the  dotted circle corresponds
to the $D_{\sun}=6\kpc$ limit. The symbols correspond to the estimated
age  of  the  open  clusters:  ${\rm  \log(age)<8.5}$  (stars),  ${\rm
8.5<\log(age)<9.0}$   (empty  triangles)  and   ${\rm  \log(age)>9.0}$
(filled triangles). All the open clusters have $|b|<18\deg$.}
\end{figure}

It is also interesting to study the  distribution of  open clusters in
light of the discovery of  the Canis Major object. Figure~13 shows the
open clusters in the WEBDA database \citep{mermilliod}
\footnote
{The WEBDA database, developed and maintained by J.-C. Mermilliod, can
be found on ${\tt http://obswww.unige.ch/webda/webda.html}$}
with $D_{\sun}>6\kpc$; a puzzling asymmetry is evident with absolutely
no clusters for the $\ell<180\deg$  part of the Milky Way. Caveats may
include  strong  uncertainties  in   distance  and/or  age  since  the
directions  are often obscured  and the  cluster population  is small.
Moreover, the presence of the Outer  Arm of the Galaxy could provide a
stronger background in the $0\deg<\ell<180\deg$ quadrant against which
it could be  more difficult to identify clusters.   On the other hand,
part of the $180\deg<\ell<270\deg$  quadrant is behind the Perseus Arm
which  is much  closer to  the  Sun: Therefore  the considered  region
suffers  from a less  severe contamination  and the  identification of
remote clusters is easier.

With these  caveats in mind, it  is still possible that  part of these
remote open clusters belong to  the Canis Major galaxy.  In particular
Dol~25         $(\ell=211.9\deg,b=-1.29\deg)$         and         H~18
$(\ell=243.1\deg,b=0.44\deg)$    which   are    both    young   (${\rm
\log(age)<8.5}$) and with a  Heliocentric distance that corresponds to
the  Canis  Major  object  could   be  the  consequence  of  a  strong
perturbation  of  the  disk  by  the accreted  dwarf  Galaxy  and  the
resulting enhanced star-formation activity  in the disk.  We also note
that the position  of the old open cluster  AM2 (previously classified
as a globular)  nearly coincides with that of the  central peak of the
Canis  Major  structure, within  the  (large) distance  uncertainties.
Unfortunately  the  lack of  any  radial  velocity  estimate for  this
cluster prevents  any definite conclusion about  its possible physical
association with the newly discovered object.

\section{N-body simulation}

We undertook a series of N-body experiments in an attempt to determine
whether a single accretion of  a dwarf galaxy could be responsible for
the detected  over-densities and the correlations  seen among globular
and open clusters. In this first study, we concentrate on constructing
a simple model  to guide interpretation of the  observations, and will
defer a more detailed analysis to a subsequent contribution.

For our  simulations, we take a  static Milky Way  potential, which is
not  perturbed by  the  passage  of the  accreted  dwarf galaxy.   The
potential model 2b of \citet{dehnen98a}  is adopted, but with the halo
component modeled by a spherical NFW halo.  Taking a static potential
is unrealistic,  of course, but it  saves a great  deal of computation
time,  making the  simulation tractable.   The key  effects  that this
approach leaves out  are the back-reaction of the  dwarf galaxy on the
Galactic  disk, and  the back-reaction  on  the halo.   The disk  will
likely  become  deformed  (though  note  that the  progenitor  of  the
Galactic ``Ring''  has an  angular momentum vector  that is  almost at
right angles to  that of the Galactic warp, so it  is unlikely to warp
the  disk strongly),  and these  deformations will  in turn  alter the
trajectory of  the accreted  object.  Since we  do not include  a live
halo component, we neglect the back-reaction on the halo (formation of
a wake), and the resulting dynamical friction on the dwarf galaxy.

A useful working picture is that  the dwarf galaxy began its life as a
typical dark matter-dominated dwarf  satellite galaxy, far away in the
Milky  Way halo.   The dark  matter halo  of the  dwarf  galaxy became
progressively more  and more disrupted as the  orbit decayed. However,
the  central  stellar  component  stayed largely  intact  during  this
initial phase.  We  start our simulation $2\Gyr$ ago,  and assume that
by  this stage  much of  the  dark matter  has been  removed from  the
system, leaving a much less  massive dwarf galaxy that will not suffer
significant  dynamical friction.   We therefore  do not  add dynamical
friction in our simulation.

The   simulations   were   performed   with   falcON,   a   fast   and
momentum-conserving tree  code \citep{dehnen00,dehnen02}.  Gravity was
softened with the  kernel ``${\tt P_2}$'', with a  softening length of
$0.1\kpc$.   The minimum  time-step was  set to  $2^{-14}  = 6.1\times
10^{-5} \Gyr$, and  the tolerance parameter $\theta =  0.6$.  In order
to obtain a starting position and velocity for the dwarf galaxy model,
we refined iteratively an initial  guess of these values, and required
that the resulting orbit fit  the velocity and position information of
the Y03  SDSS and  the 2MASS  fields presented above,  as well  as the
globular  clusters.   Acceptable  orbits  could  be  fit  for  both  a
co-rotating and  a counter-rotating object,  so we examine  both cases
below.  These fitted orbits have an azimuthal period of $0.45\Gyr$ and
$0.3\Gyr$,  respectively,  for  the co-rotating  and  counter-rotating
models.

The initial  dwarf galaxy was modeled  with a King  model with $10^5$
particles.   We  experimented with  several  such  models, trying  out
masses  in the  range  of  $2\times 10^8$  to  $10^9\msun$ (the  range
suggested by \citealt{ibata03}, and  which is consistent with the mass
estimate  derived   above),  and   with  different  tidal   radii  and
concentration  parameters.  Some  fine-tuning of  these  parameters is
needed  to  give both  a  stream  and  remnant that  approximates  the
observed distribution.   One of the  better fitting models had  a mass
$5\times10^8\msun$,  tidal radius  $2.5\kpc$ and  $W0=4.5$, and  it is
this model that we will discuss below. However, these values are by no
means the  required initial parameters  of the satellite  galaxy. They
are sensitive  to the orbit, and  must therefore also  be sensitive to
the decay  rate of the satellite  due to dynamical  friction, which we
have neglected. The model is therefore only illustrative.

In  the top  panels of  Figures~14 and  15, we  show  respectively the
$x$--$y$ configuration (looking down onto the Milky Way from the north
Galactic pole) at  the end of the simulation,  for the co-rotating and
counter-rotating models.  Given the  simplicity of the assumptions the
similarity  between the  simulations and  observations  is remarkable,
though  we feel  that  a  formal statistical  comparison  is not  very
meaningful at  this stage due  to the fact  that we have left  out the
self-gravity of the Milky Way.   However, what the simulations do show
is that the  streams that arise naturally during  the tidal disruption
of a  dwarf galaxy can be pulled  out to be seen  as multiple streams,
even in  the relatively short  time ($\sim 2\Gyr$) that  we simulated.
This  may account  for the  different  peaks in  the distance  modulus
distributions  (Figures~2 and  8).  Figures~14  and 15  also  show the
positions of the globular  clusters that have Galactocentric distances
$R_{GC} > 8\kpc$,  and that lie within $8\kpc$  of the Galactic plane.
To  select the  closest star  clusters  to our  simulation, we  choose
somewhat arbitrarily a simple statistic:
$ \chi^2 = \big( {{|\vec{x}-\vec{x}_{model}|}\over{2\kpc}} \big)^2  
         + \big( {{v_\odot - v_{\odot \, model}}\over{20\kms}} \big)^2 $;
we list  the globular and  open clusters that  have $\chi < 3$  in the
prograde and  retrograde models, in  Tables~1 and 2  respectively.  We
defer analyzing the likelihood of the alignment of these star clusters
with the Canis Major stream to a future contribution when we present a
more realistic simulation with a live Milky Way.

The distribution of the simulation particles projected onto the sky is
shown in Figure~16; the  models nicely reproduce the curving behaviour
of the Northern Arc  observed towards the Galactic anticentre (compare
to   Figure~4),  the  Canis   Major  over-density   as  well   as  the
over-densities   ``A''   and  ``B''   (see   Figure~5)  seen   towards
$\ell=70\deg$  and $\ell=300\deg$.   The Heliocentric  radial velocity
profiles of  the models  are given in  Figure~17. To our  surprize, we
find that current data do not appear to prefer the prograde model over
the  retrograde.  Future measurements  of the  radial velocity  of the
stream  in   directions  sufficiently  far  away   from  the  Galactic
anticentre should resolve this uncertainty.

The numerical  simulations presented here  lend support to  the notion
that the Canis Major over-density  is the remnant of a larger accreted
galaxy  whose tidal  disruption gave  rise to  the asymmetries  in the
Galactic  M-giant distributions that  we report  above.  If  we accept
this model, we find that the  progenitor had an orbit that was closely
aligned with the  Galactic Plane (the orbit is  inclined at only $\sim
10\deg$ from the  Plane). It is unclear at  present whether the object
follows a prograde or a retrograde orbit, though a-priori the prograde
case would appear to be more  likely since it would allow for stronger
interactions with the Galactic disk, due to which the orbit would have
decayed to its present,  rather circular, state.  The remaining debris
has an apocentre of $\sim 20\kpc$  and a pericentre close to the Solar
radius. Indeed,  according to the simulations  displayed in Figures~14
and  15, the debris  from the  Canis Major  dwarf streams  through the
Solar  Neighbourhood.  Comparing  the  kinematics of  the Canis  Major
stream  to the stream  discovered by  \citet{helmi99}, the  only other
large stream known to pass close to the Sun, shows that the streams of
our models have very different angular momentum.  The angular momentum
distribution   of  the   particles   within  $2\kpc$   of  the   Solar
Neighbourhood  is shown  in  Figure~18; these  particles display  very
different  kinematics,  and  are  unlikely  to be  associated  to  the
\citet{helmi99} stream.

\begin{figure}
\ifthenelse{\UseFigs=1}{
\includegraphics[angle=0,bb= 115 170 460 700,clip,width=\hsize]{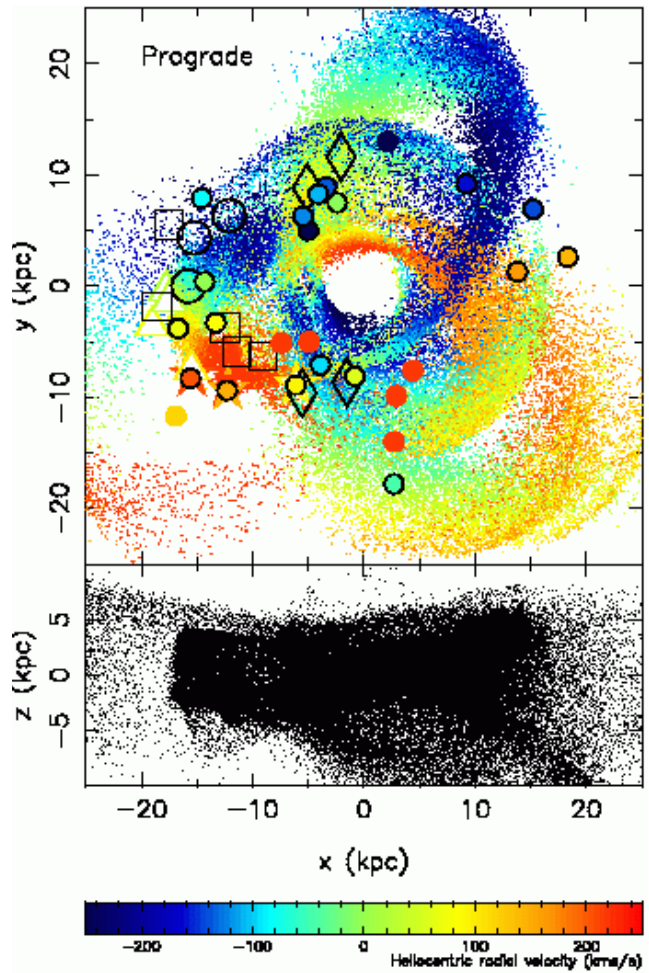}
}{}
\caption{Two  different projections  of  an N-body  simulation of  the
disruption of  the progenitor of  the Galactic ``Ring'' and  the Canis
Major structure.  The Galactic Centre is at the origin, and the Sun at
($x,y,z=-8,0,0\kpc$). The dwarf galaxy  model was placed on an
orbit that co-rotates  with the Galactic disk.  The  upper panel shows
the  $x$--$y$ distribution  (i.e.   as seen  from  the Galactic  north
pole),  with colour-coded  radial velocities.  In black,  we  show the
positions and distances of  the INT fields from \citet{ibata03} (large
open circles), and the 2MASS data presented in this contribution (open
squares and diamonds).  The  diamonds represent those fields for which
a distance  estimate was secured by summing  in Galactocentric, rather
than Heliocentric  distance.  The colour-coded graph  markers show the
positions,  distances and  velocities  of the  four  SDSS fields  from
\citet{yanny}  (open  triangles),  and  of  the  4  globular  clusters
detailed in  section~4 (star  symbols).  The filled  circles represent
all globular and open clusters of known radial velocity within $8\kpc$
of the Galactic plane,  and that have Galactocentric distances $R_{GC}
> 8\kpc$; the subset of these that are circled black are those objects
listed in  Table~1, which are  close in phase-space to  the simulation
particles.  This velocity-distance-position coincidence with the model
suggests that  several globular clusters  around the Milky Way  may be
associated with the Canis Major accretion event.}
\end{figure}

\begin{figure}
\ifthenelse{\UseFigs=1}{
\includegraphics[angle=0,bb= 115 170 460 700,clip,width=\hsize]{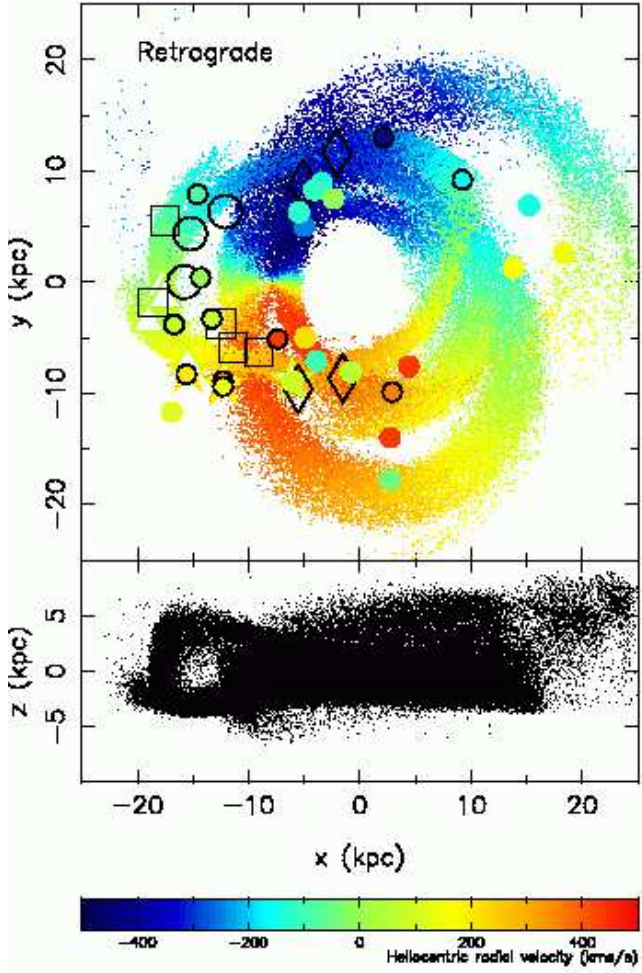}
}{}
\caption{As Figure~14, but for a model on a retrograde orbit.}
\end{figure}

\begin{figure}
\ifthenelse{\UseFigs=1}{
\includegraphics[angle=0,bb= 30 220 540 540,clip,width=\hsize]{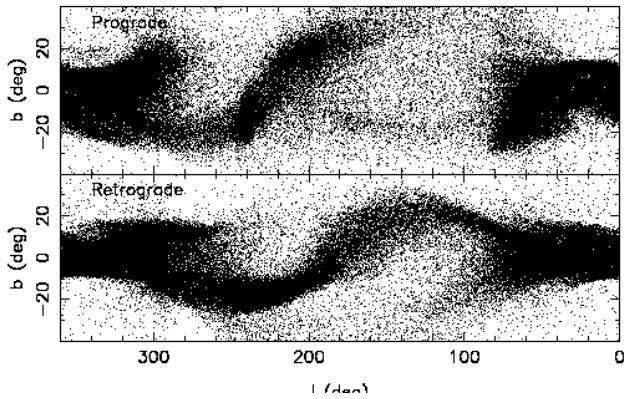}
}{}
\caption{The  $\ell$--$b$ distribution  of the  prograde  model (upper
panel)  and retrograde  model  (lower panel).   Most  of the  observed
large-scale  asymmetries  of  the  Galactic M-giant  distribution  are
reproduced  in this projection  of the  simulations: the  broad curved
distribution  of the  Northern Arc  seen towards  the  anticentre, the
strong over-density seen towards  Canis Major at $\ell \sim 240^\circ,
b \sim -10^\circ$,  a weak feature similar to  the ``Southern Arc'' is
also present,  plus the over-densities  ``A'' and ``B'' at  $\Phi \sim
100^\circ$, and $\Phi \sim 260^\circ$.}
\end{figure}

\begin{figure}
\ifthenelse{\UseFigs=1}{
\includegraphics[angle=0,bb= 30 40 540 710,clip,width=\hsize]{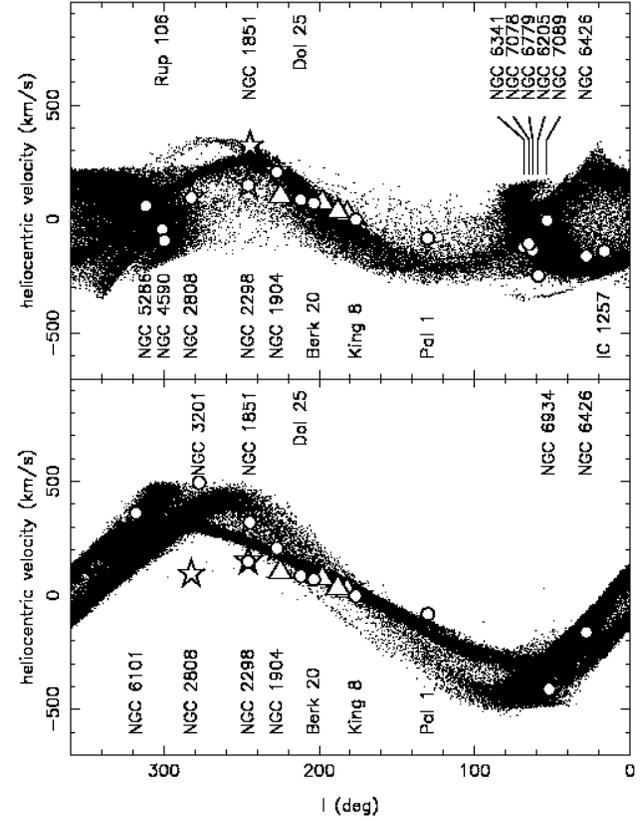}
}{}
\caption{  The  Heliocentric radial  velocity  of  the prograde  model
(upper  panel) and  retrograde model  (lower panel)  as a  function of
Galactic  longitude,  is  compared   to  the  Y03  SDSS  fields  (open
triangles) and  the 4 globular clusters  NGC~1851, NGC~1904, NGC~2298,
and NGC~2808 (star symbols). In  addition, the open circles show, from
right to left, the positions of  the star clusters in Table~1 (for the
top panel) and Table~2 (for the bottom panel), in the order listed.}
\end{figure}

\begin{figure}
\ifthenelse{\UseFigs=1}{
\includegraphics[angle=0,width=\hsize]{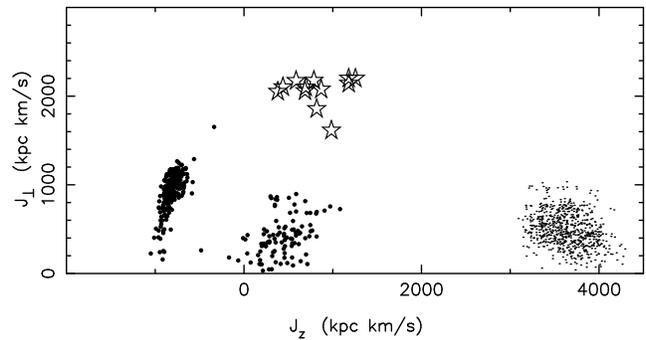}
}{}
\caption{The  larger dots  show the  angular momentum  distribution of
particles in the prograde simulation  within $2\kpc$ of the Sun, while
the  smaller  dots show  the  same  parameters  for particles  in  the
retrograde simulation.  These  distributions differ substantially from
that  found by  \citet{helmi99} for  the star  stream detected  in the
Solar Neighbourhood  (star symbols).  The particles in  both the Canis
Major  simulations  follow a  more  planar  orbit,  with the  prograde
particles  having  typical   Galactic  disk  kinematics.   This  large
difference implies that the Canis  Major stream is not associated with
the \citet{helmi99} structure.}
\end{figure}

\section{Discussion and Conclusions}

We have shown  that the distribution of M-giant  stars at low latitude
is highly  asymmetric.  The  most striking over-density  is due  to an
elliptical object present  in the disk, in the  direction of the Canis
Major   constellation   in  the   southern   Galactic  hemisphere   at
$D_{GC}=12.0\pm1.2\kpc$. This structure contains as many M-giant stars
as  the  remnant  of  the  Sagittarius dwarf  galaxy.   A  less  dense
structure, but of huge proportions, the Northern Arc curves around the
Galaxy in the  direction of the Galactic anticentre.   This arc ranges
from the Galactic plane at $\ell  \sim 230\deg$ to $b \sim +30\deg$ at
$\ell  \sim 140\deg$.   Between $140\deg<\ell<180\deg$  it has  a mean
displacement of $\sim20\deg$ above the Galactic Plane, with a distance
of  $D_{GC}=18.1\pm0.1\kpc$, it  is  separated from  the  disk in  the
radial direction along  the line of sight, and  has a Gaussian profile
of width $11.6\deg\pm1.9\deg$.

Several  other   structures  are  also  detected.    In  the  southern
hemisphere,  the  ``Southern   Arc''  population  with  Galactocentric
distance  $12.0\kpc<D_{GC}<15.1\kpc$ is  observed  between $110\deg  <
\ell  <  210\deg$;  this  may  be the  continuation  of  the  southern
hemisphere  structure  reported  by  Y03 for  $R_{GC}=20\pm2\kpc$  and
$180\deg  < \ell <  227\deg$.  Two  other over-densities  are detected
towards the  Galactic bulge, and close  to the Galactic  Plane: one is
found between  $230\deg < \Phi < 270\deg$  at $D_{GC}<14.5\kpc$, while
the other is between $90\deg < \Phi < 130\deg$ at $D_{GC}=10.5\kpc$.

Given these observations, we examine the idea of an in-plane accretion
of a dwarf satellite galaxy  onto the Milky Way.  The realization that
the four globular clusters NGC~1851, NGC~1904, NGC~2298, and NGC~2808,
are close to each other in  phase space and are also close in position
and distance to the Canis Major over-density also lends strong support
to this possibility.  In this scenario the remnant of the satellite is
likely the object  in the direction of Canis  Major.  During the tidal
disruption  process, the  dwarf galaxy  loses stars  in a  long stream
which loops  around the  Milky Way  and is now  seen as  the Monoceros
``Ring'' structure.  The two other  hypotheses presented by I03 do not
seem to  be allowed  by our observations.   Indeed, the height  of the
structure (generally more than $3\kpc$) could hardly be explained by a
spiral  arm.    Furthermore,  the  multiple  peaks   in  the  distance
distribution  are more suggestive  of multiple  crossings of  a stream
rather  than  of  a  ring-like  Galactic structure,  produced  by  the
dynamical evolution of ancient stellar warps.

The simple N-body  models we presented reproduce many  of the observed
spatial  features, including  the Canis  Major over-density,  the vast
ring-like  arc  towards the  Galactic  anticentre, the  over-densities
towards  $\ell=70\deg$ and  $\ell=290\deg$,  as well  as faint  debris
scattered  out to  large  distance  below the  Galactic  Plane in  the
anticentre  direction.   The   models  also  reproduce  the  available
kinematic  data  and  point  to  the  possibility  that  several  star
clusters,  apart from  those close  to the  Canis  Major over-density,
could  be related  to the  Canis Major  Stream.  The  success  of this
simple  model  strongly  suggests  that  many,  if  not  all,  of  the
asymmetries in  the Galactic  M-giant population presented  above, are
due to a single accretion  event. The accreted satellite must have had
an orbital plane closely aligned to the plane of the Milky Way galaxy.
Our analysis  shows that the ``shell'' model  of \citet{helmi03} gives
poor agreement with the observations,  but we confirm the tidal stream
model for the ``Monoceros Ring'' proposed by Y03.

The observations and  model also suggest a connection  to the Galactic
thick disk.  The orbital eccentricity of  the model is $e = (r_{apo} -
r_{peri})/(r_{apo} + r_{peri}) \sim 0.5$,  a value that is typical for
Galactic  thick disk  stars of  metallicity ${\rm  [Fe/H]  \sim -1.2}$
\citep{chiba}.  Furthermore, the  vertical exponential scale height of
the debris $0.73\pm0.05\kpc$, is very close to the thick disk value of
$0.8\kpc$ determined  by \citet{reyle}. It is  therefore possible that
we have uncovered  one of the galactic building  blocks from which the
thick  disk grows.   This analysis  suggests  that the  thick disk  is
continually growing,  even up to the present  time, through accretions
of dwarf galaxies in co-planar orbits. Although with a progenitor mass
of $\sim 10^9\msun$,  as suggested by our analysis,  only a handful of
accretions of  the magnitude of that  of the Canis  Major dwarf galaxy
are  needed to populate  the thick  disk entirely.   Globular clusters
with disk-like kinematics may also  have been brought to their present
locations  in this  way. The  observations and  modeling  we presented
strongly  support the  analysis  of \citet{abadi},  in  which a  large
proportion  of  thick  stars  have  their origin  in  accreted  galaxy
fragments.  If the  scenario we have developed here  is correct, thick
disks should  be a generic  component of giant spiral  galaxies, since
accretions are a generic  feature of galaxy formation. Furthermore, at
the  edge of the  disk where  the dynamical  times are  longest, their
structure will  likely be a  complex mess where streaming  debris from
the  most  recent  accretions  piles  up at  apocentre.   These  messy
structures  may indeed  already have  been detected  in  the Andromeda
galaxy \citep{ibata01b,ferguson02}.

Further observational  work is needed  to constrain the  kinematics of
the  structures  we report  here.   These  constraints  are needed  to
determine  whether the structure  follows a  prograde or  a retrograde
orbit, to verify  the kinematic predictions of our  models, as well as
to  guide future simulations.   Detailed abundance  distributions will
also be invaluable  in order to study the  chemical difference between
an accreted  population, such as that  of the Canis  Major galaxy, and
that  of the  normal Galactic  disk.  Future  simulations with  a live
Milky   Way  will  provide   a  more   realistic  comparison   to  the
observations. It  will be interesting  to investigate the  reaction of
the Galactic disk stars and in  particular the reaction of the HI gas,
to  this sizable  accretion, while  the back-reaction  onto  the Canis
Major dwarf galaxy  will be essential in order  to model its dynamical
evolution accurately.

\section*{Acknowledgments}

We are grateful to A.  Bragaglia and L.  Monaco for useful suggestions
and discussion, and  to A. Helmi for kindly  providing the kinematics
of  the  Solar  Neighbourhood   stream  stars.   MB  acknowledges  the
financial  support   to  this  research  by   the  Italian  Ministero
dell'Universit\`a  e  della Ricerca  Scientifica  (MURST) through  the
grant p. 2001028879, assigned to the project {\em Origin and Evolution
of the Galactic Spheroid}.

This publication  makes use of data  products from the  Two Micron All
Sky  Survey,   which  is  a   joint  project  of  the   University  of
Massachusetts    and   the    Infrared    Processing   and    Analysis
Center/California  Institute  of Technology,  funded  by the  National
Aeronautics  and   Space  Administration  and   the  National  Science
Foundation.

\begin{table}
\begin{center}
\caption{Star clusters close in phase-space to the prograde model.}
\begin{tabular}{lccccc}
\hline\hline
Name      & $\ell$ & $|\vec{x}-\vec{x}_{model}|$  & $v_\odot$ & $v_{\odot
\,  model} $
 & $\chi$ \\
          & (deg) &  (kpc) & $(\kms)$  & $(\kms)$ &  \\
\hline
IC 1257  &    16.5 &    2.7 & -140.2 & -138.8 &    1.8 \\
NGC 6426 &    28.1 &    2.2 & -162.0 & -148.3 &    1.7 \\
NGC 7089 &    53.4 &    1.5 &   -5.3 &   15.8 &    1.6 \\
NGC 6205 &    59.0 &    2.7 & -246.6 & -235.7 &    2.2 \\
NGC 6779 &    62.7 &    1.0 & -135.7 & -137.1 &    0.2 \\
NGC 7078 &    65.0 &    2.2 & -107.3 &  -97.3 &    1.5 \\
NGC 6341 &    68.3 &    0.9 & -122.2 & -114.3 &    0.4 \\
Pal 1    &   130.1 &    2.6 &  -82.8 &  -88.0 &    1.8 \\
King 8   &   176.4 &    0.9 &   -2.0 &   -9.9 &    0.4 \\
Berk 20  &   203.5 &    0.8 &   70.0 &   69.6 &    0.1 \\
Dol 25   &   211.9 &    1.1 &   85.0 &   73.8 &    0.6 \\
NGC 1904 &   227.2 &    2.8 &  206.0 &  195.3 &    2.2 \\
NGC 2298 &   245.6 &    3.3 &  148.9 &  150.9 &    2.7 \\
NGC 2808 &   282.2 &    1.4 &   93.6 &   91.2 &    0.5 \\
NGC 4590 &   299.6 &    2.5 &  -94.3 & -102.0 &    1.7 \\
Rup 106  &   300.9 &    1.6 &  -44.0 &  -33.2 &    0.9 \\
NGC 5286 &   311.6 &    2.7 &   57.4 &   63.4 &    1.9 \\
\hline
\end{tabular}
\end{center}
Data extracted  from \citet{harris} and  \citet{mermilliod}.  Column 2
states  the  Galactic  longitude   of  the  cluster;  column  3  lists
$|\vec{x}-\vec{x}_{model}|$,  the difference  in position  between the
simulation and  the cluster; columns  4 and 5 list,  respectively, the
Heliocentric radial  velocity of the cluster  and simulation; finally,
in column 6, we give the value of the $\chi$ closeness statistic.
\end{table}

\begin{table}
\begin{center}
\caption{Star clusters close in phase-space to the retrograde model 
(data arranged as in Table~1).}
\begin{tabular}{lccccc}
\hline\hline
Name      & $\ell$ & $|\vec{x}-\vec{x}_{model}|$  & $v_\odot$ & $v_{\odot
\,  model} $
 & $\chi$ \\
          & (deg) &  (kpc) & $(\kms)$  & $(\kms)$ &  \\
\hline
NGC 6426 &    28.1 &    1.6 & -162.0 & -169.0 &    0.7 \\
NGC 6934 &    52.1 &    3.2 & -411.4 & -411.2 &    2.5 \\
Pal 1    &   130.1 &    1.1 &  -82.8 &  -92.8 &    0.5 \\
King 8   &   176.4 &    1.8 &   -2.0 &   -9.3 &    0.9 \\
Berk 20  &   203.5 &    0.6 &   70.0 &   71.4 &    0.1 \\
Dol 25   &   211.9 &    1.9 &   85.0 &   87.8 &    0.9 \\
NGC 1904 &   227.2 &    2.8 &  206.0 &  201.1 &    2.0 \\
NGC 1851 &   244.5 &    1.6 &  320.5 &  340.4 &    1.7 \\
NGC 2298 &   245.6 &    1.9 &  148.9 &  150.7 &    0.9 \\
NGC 3201 &   277.2 &    1.2 &  494.0 &  482.5 &    0.7 \\
NGC 6101 &   317.8 &    0.5 &  361.4 &  361.1 &    0.1 \\
\hline
\end{tabular}
\end{center}
\end{table}

\newcommand{\mnras}{MNRAS}
\newcommand{\nat}{Nature}
\newcommand{\araa}{ARAA}
\newcommand{\aj}{AJ}
\newcommand{\apj}{ApJ}
\newcommand{\apjl}{ApJ}
\newcommand{\apjs}{ApJSupp}
\newcommand{\aap}{A\&A}
\newcommand{\aaps}{A\&ASupp}
\newcommand{\pasp}{PASP}

\end{document}